\documentclass[aps,prc,superscriptaddress,showpacs,nofootinbib,floatfix,twocolumn]{revtex4}

\usepackage{doi,graphicx,amssymb,amsmath,dcolumn,gensymb}

\newcommand{\unit}[1]{\ensuremath{\, \mathrm{#1}}}
\newcommand{\etal}{\textit{et al.}}

\newcolumntype{d}[1]{D{.}{.}{#1} }
\usepackage[percent]{overpic}

\begin{document}

\title{Effect of topological defects on ``nuclear pasta'' observables}
\author{A. S. Schneider}\email{andschn@caltech.edu}
\affiliation{TAPIR, California Institute of Technology, Pasadena, CA 91125, USA}
\author{D. K. Berry}\email{dkberry@iu.edu}
\author{M. E. Caplan}\email{mecaplan@indiana.edu}
\author{C. J. Horowitz}\email{horowit@indiana.edu}
\author{Z. Lin}\email{zidulin@imail.iu.edu}
\affiliation{Center for Exploration of Energy and Matter and Department of Physics, Indiana University, Bloomington, IN 47405, USA}
\date{\today}
\begin{abstract}

\begin{description}
\item[Background]
The ``pasta'' phase of nuclear matter may play an important role in the structure and evolution of neutron stars. Recent works suggest nuclear pasta has a high resistivity which could be explained by the presence of long lived topological defects. The defects act as impurities that decrease thermal and electrical conductivity of the pasta.
\item[Purpose]
To quantify how topological defects affect transport properties of nuclear pasta and estimate this effect using an impurity parameter $Q_{\text{imp}}$.
\item[Methods]
Contrast molecular dynamics simulations of up to 409\,600 nucleons arranged in parallel nuclear pasta slabs (perfect pasta) with simulations of pasta slabs connected by topological defects (impure pasta). 
From these simulations compare the viscosity and heat conductivity of perfect and impure pasta to obtain an effective impurity parameter $Q_{\text{imp}}$ due to the presence of defects.
\item[Results]
Both the viscosity and thermal conductivity calculated for both perfect and impure pasta are anisotropic, peaking along directions perpendicular to the slabs and reaching a minimum close to zero parallel to them. In our 409\,600 nucleon simulation topological defects connecting slabs of pasta reduce both the thermal conductivity and viscosity on average by about 37\%. We estimate an effective impurity parameter due to the defects of order $Q_{\text{imp}}\sim30$.
\item[Conclusions] 
Topological defects in the pasta phase of nuclear matter have an effect similar to impurities in a crystal lattice.  The irregularities introduced by the defects reduce the thermal and electrical conductivities and the viscosity of the system.  This effect can be parameterized by a large impurity parameter $Q_{\text{imp}}\sim30$.
\end{description}

\end{abstract}



\maketitle

\section{Introduction}\label{sec:Intro}

A major goal of nuclear physics is to determine the equation of state (EoS) of dense matter. Often the EoS is parametrized in terms of a few observables such as the binding and symmetry energies, stiffness and compressibility of nuclear matter; all of which can be constrained by experiments and astrophysical observations \cite{0004-637X-771-1-51,0004-637X-773-1-11}. From the EoS one may determine properties of nuclei such as radii, binding energies and neutron skin thickness \cite{PhysRevLett.86.5647} as well as mass-radius relation of neutron stars (NSs) and their maximum mass \cite{2041-8205-765-1-L5}. Some of these observables are correlated and, thus, measurement of one observable places constraints on the range of another. For instance, Fattoyev and Piekarewicz showed that the skin thickness of $^{208}$Pb is significantly correlated to the radii of a $1.0M_\odot$ NS, but not to NSs maximum mass \cite{PhysRevC.84.064302}.

Besides the EoS, transport properties of dense matter are another key factor determining the structure and evolution of a NS. For example, shear viscosity regulates dissipation of hydrodynamical motions such as instabilities associated with the emission of gravitational waves \cite{doi:10.1142/S0218271801001062} while thermal conductivity determines the temperature profile and evolution of the star \cite{Lim:2015lia}. Hence, stringent limits on some EoS parameters as well as the inner structure and transport properties of NSs can be placed from detection of gravitational waves \cite{PhysRevD.79.124033,PhysRevLett.80.4843}, measurements of surface temperatures and luminosities of isolated NS \cite{Lattimer2007109,PhysRevC.92.012801} and from the cooling curves of accreting NSs during quiescence \cite{0004-637X-774-2-131,2041-8205-722-2-L137}. In this work we are specifically interested in the transport properties of the inner crust of a NS, where matter is believed to arrange itself into exotic neutron-rich nonspherical shapes known as \textit{nuclear pasta}  \cite{Pethick19987,PhysRevLett.50.2066}. Though nuclear pasta is not expected to significantly alter the EoS of dense matter it does have a large effect on its transport properties \cite{PhysRevC.69.045804,PhysRevC.78.035806,Yakovlev11102015}.

In a recent study Pons \etal\, suggested that existing limits on the maximum period of isolated x-ray pulsars could be a consequence of a highly resistive layer present in the crust-core boundary (inner crust) of neutron stars \cite{pons2013highly}. This resistive layer may be accounted for by the existence of a disordered phase or ``impurities'' in the inner crust of the NS, such as a pasta phase. In a recent work we suggested that topological defects in the pasta phase could act as impurities that decrease both thermal and electrical conductivities in the NS inner crust \cite{PhysRevLett.114.031102}. In that work we made a qualitative estimate of the impurity parameter due to the topological defects and obtained its effect on the cooling curve of the low mass x-ray binary LMXB 1659-29. In this work we aim to quantify the effective impurity parameter and determine its effect on the viscosity and thermal conductivity of NSs inner crust. For studies on the thermal conductivity of the outer crust of neutron stars, where ions are the relevant degrees of freedom, see References \cite{PhysRevE.79.026103,PhysRevC.92.045809,Roggero:2016fev}.

The paper is outlined as follows. In Section \ref{sec:Formalism} we introduce our molecular dynamics formalism and discuss how to obtain the structure factor of the pasta and observables related to it. In Section \ref{sec:Results} we show our results and discuss an estimate for the impurity parameter due to topological defects on the pasta structure. We conclude in Section \ref{sec:Conclusions}.

\section{Formalism}\label{sec:Formalism}

We use the same molecular dynamics (MD) formalism of our previous works, for an example see Reference \cite{PhysRevC.88.065807}. We assume nucleons are point-like particles immersed in a background electron gas and interact via a central two-body potential that depends on their isospins: 
\begin{subequations}\label{eq:potential}
\begin{align}
 V_{np}(r)&=a e^{-r^2/\Lambda}+[b-c]e^{-r^2/2\Lambda}\\
 V_{nn}(r)&=a e^{-r^2/\Lambda}+[b+c]e^{-r^2/2\Lambda}\\
 V_{pp}(r)&=a e^{-r^2/\Lambda}+[b+c]e^{-r^2/2\Lambda}+\frac{\alpha}{r}e^{-r/\lambda}.
\end{align}
\end{subequations}
Here, $n$ $(p)$ denotes neutrons (protons) and $r$ the distance between a nucleon pair. The parameters $a$, $b$, $c$ and $\Lambda$ given in Table \ref{Tab:parameters} are fit to reproduce some properties of finite nuclei, pure neutron matter and symmetrical nuclear matter \cite{PhysRevC.69.045804}. While interactions that involve neutrons only contain short-range terms, proton-proton interactions involve the long-range Coulomb repulsion screened by the background electron gas. As in previous works we set the electron screening length $\lambda$ to 10\unit{fm} \cite{PhysRevLett.114.031102}. For the simulations performed in this work, see Section \ref{sec:Results}, this value is slightly smaller than 
the one obtained considering non-interacting relativistic electrons, $\lambda=12.35\unit{fm}$. 
We do not expect this difference to influence our results significantly \cite{PhysRevC.90.055805}.

\begin{table}[h]
\caption{\label{Tab:parameters} Nuclear interaction parameters.}
\begin{ruledtabular}
\begin{tabular}{*{4}{c}}
$a$ (MeV) &$b$ (MeV)&$c$ (MeV)&$\Lambda$ (fm$^{2}$) \\
\hline
  110     &  $-$26    &   24    &    1.25      \\
\end{tabular}
\end{ruledtabular}
\end{table}

In our simulations we consider cubic volumes with periodic boundary conditions and assume that each nucleon interacts solely with the nearest periodic image of other nucleons. All simulations were run on the Big Red 2 supercomputer of Indiana University using the IUMD CPU/GPU hybrid code detailed in References \cite{PhysRevC.90.055805,PhysRevC.91.065802}.

For each simulation we obtain a trajectory file and use it to determine the structure factor for protons and neutrons following the recipe of References \cite{PhysRevC.78.035806,PhysRevC.90.055805}. The structure factor $S_i(\boldsymbol{q})$ for nucleons of type $i=n,\,p$ is given by the time average of the nucleon density in momentum space:
\begin{equation}\label{eq:sq}
 S_i(\boldsymbol{q})=\langle\rho_i^*(\boldsymbol{q},t)\rho_i(\boldsymbol{q},t)\rangle_t
                    -\langle\rho_i^*(\boldsymbol{q},t)\rangle_t\langle\rho_i(\boldsymbol{q},t)\rangle_t
\end{equation}
where $\rho_i(\boldsymbol{q},t)=N_i^{-1/2}\sum_{j=1}^{N_i}e^{i\boldsymbol{q}\cdot\boldsymbol{r}_j(t)}$ 
is the nucleon density in momentum space, $N_i$ the number of nucleons of type $i$, $\boldsymbol{r}_j(t)$ the position of the $j$-th nucleon of type $i$ at time $t$ and the angled brackets $\langle{A}\rangle_a$ denote the average of quantity $A$ over a set of $a$. In Reference \cite{PhysRevC.78.035806} Horowitz and Berry used an angle average of the proton structure factor, 
$S_p(q)=\langle S_p(\boldsymbol{q})\rangle_{\vert\boldsymbol{q}\vert}$, to calculate pasta observables.

Once the proton structure factor has been obtained it may be used to calculate the shear viscosity $\eta$, electron conductivity $\sigma$ and thermal conductivity $\kappa$. Using the results of References \cite{chugunov2005,Nandkumar01081984} and approximations of Reference \cite{PhysRevC.78.035806} one obtains for the pasta phase that
\begin{equation}\label{eq:visc1}
 \eta=\frac{\pi v^2_Fn_e}{20\alpha^2\Lambda^\eta_{\text ep}},
\end{equation}
\begin{equation}\label{eq:elec1}
 \sigma=\frac{v^2_F k_F}{4\pi\alpha\Lambda^\sigma_{\text ep}}
\end{equation}
and
\begin{equation}\label{eq:heat1}
 \kappa=\frac{\pi v^2_F k_F k_B^2 T}{12\alpha^2\Lambda^\kappa_{\text ep}}.
\end{equation}
Here, $v_F$ and $k_F$ are the Fermi velocity and momentum of electrons, $n_e$ the electron density, $\alpha$ the fine structure constant, $k_B$ is Boltzmann's constant and $T$ the temperature of the system. The Coulomb logarithms $\Lambda^i_{ep}$ for $i=\eta$, $\kappa$ and $\sigma$ are approximated 
by the following integrals \cite{PhysRevC.78.035806}
\begin{equation}\label{eq:visc2}
 \Lambda^\eta_{\text ep}=\int_0^{2k_F}\frac{dq}{q\epsilon^2(q)}\left(1-\frac{q^2}{4k_F^2}\right)\left(1-\frac{v_F^2q^2}{4k_F^2}\right)S_p(q)
\end{equation}
and 
\begin{equation}\label{eq:heat2}
 \Lambda^\kappa_{\text ep}=\Lambda^\sigma_{\text ep}=\int_0^{2k_F}\frac{dq}{q\epsilon^2(q)}\left(1-\frac{v_F^2q^2}{4k_F^2}\right)S_p(q).
\end{equation}
In the equations above $\epsilon(q)=1+k_{TF}^2/q^2$ is the Thomas-Fermi approximation to the dielectric function where $k_{TF}=2k_F\sqrt{\alpha/\pi}=\lambda^{-1}$ is the inverse screening length and $k_F=(3\pi^2n_e)^{1/3}$ where $n_e$ is the electron density assuming a neutral system. As mentioned before, we used the approximation $\lambda=10\unit{fm}$, a value not far from the true screening length value $\lambda=12.35\unit{fm}$ and not expected to contribute significantly to our final results \cite{PhysRevC.90.055805}. As discussed in References \cite{Nandkumar01081984,PhysRevC.78.035806} these results are valid in 
the classical limit where excitations energies $\omega$ of the system are much smaller than its temperature $T$, \textit{ì.e.}, $\omega\ll T$. The same equations can be obtained in the zero magnetic field limit, $\boldsymbol{B}\rightarrow0$, of results derived by Yakovlev in Reference \cite{Yakovlev11102015}.

In Reference \cite{PhysRevC.78.035806} Horowitz and Berry argued that if the structure factor $S_p(q)$ can be approximated by a triangle of height $Z^*$ centered at $q^*$ with width at half maximum $\Delta q^*$ then
\begin{equation}\label{eq:Log}
 \Lambda_{ep}\approx\frac{\Delta q^* Z^*}{q^*}
\end{equation}
and the expressions for $\eta$ and $\kappa$ simplify considerably. In this expression $Z^*$ can be assumed as an effective charge of the pasta, \textit{i.e.}, the effective number of protons an electron can scatter from.

\section{Results}\label{sec:Results}

The main goal of this work is to quantify how some types of defects in the pasta topology affect its properties. We, thus, compare observables obtained for pasta shapes with no topological defects (perfect pasta) 
to the ones of pasta with topological defects (impure pasta). As in Reference \cite{PhysRevLett.114.031102} we study systems of nucleons with density $n=0.050\unit{fm}^{-3}$, proton fraction $Y_p=0.40$ at temperature $T=1\unit{MeV}$.

\subsection{Topology}\label{ssec:Topology}

\begin{figure*}[!h]
\centering
\includegraphics[width=1.0\textwidth]{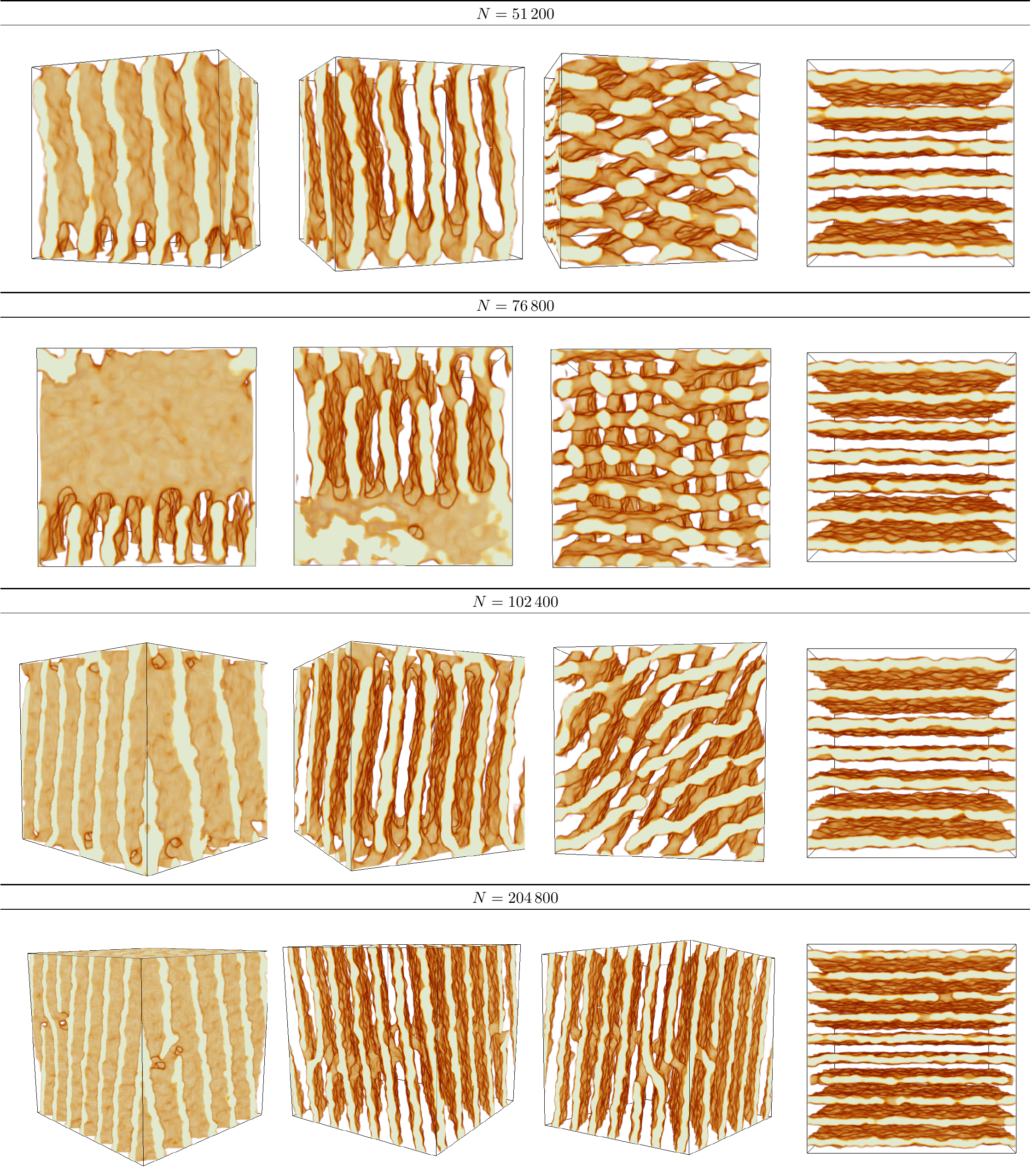}
\caption{\label{fig:configurations} (Color on line) Charge density isosurfaces of runs with 51\,200, 76\,800, 102\,400 and 204\,800 nucleons for systems with mean density $n=0.05\unit{fm}^{-3}$, temperature $kT=1.00$ MeV and proton fraction $Y_p = 0.40$. The system with 409\,600 nucleons produced defects very similar to the 204\,800 nucleon simulation and is not shown. The first three columns show the same impure pasta configuration seen from three different angles that facilitate the view of the defects. The last column shows the final configuration of the perfect pasta simulation. Golden surfaces represent isosurfaces of charge density $n_{\rm{ch}}=0.03\unit{fm}^{-3}$, while the cream color shows regions such that $n_{\rm{ch}}>0.03 \unit{fm}^{-3}$. The figures were generated with ParaView software \cite{Paraview}.}
\end{figure*}

As shown in References \cite{PhysRevC.91.065802,PhysRevLett.114.031102} the pasta phases obtained from molecular dynamics (MD) simulations started from random initial configurations at the densities, temperatures and proton fractions considered here are flat sheets connected by topological defects (impure pasta). In this work we simulate systems with 51\,200, 76\,800, 102\,400, 204\,800 and 409\,600 nucleons for at least $5\times10^6$ MD time steps of $2\unit{fm}/c$. As shown in the first three columns of Figure \ref{fig:configurations} we see that the types of defects are sensitive to the simulation size.

Following our previous works we quantify the shapes of the pasta phases using Minkowski functionals focusing on the average mean and average Gaussian curvatures: $B/A$ and $\chi/A$, respectively \cite{PhysRevC.88.065807,PhysRevC.90.055805}.

Negative values of $\chi/A$ indicate an interconnected system \cite{PhysRevC.91.025801}. Since our systems are mainly made of flat sheets connected by topological defects the magnitude of the curvatures serves as an indication of defect density. In Reference \cite{Horowitz:2015gda} we discuss how the geometry of defects may be determined by a Helfrinch-Canham type Hamiltonian \cite{citeulike:6130970,Canham197061} that involves both curvatures.

\begin{figure}[!h]
\centering
\begin{overpic}[trim= -10 0 0 0,clip,width=0.495\textwidth]{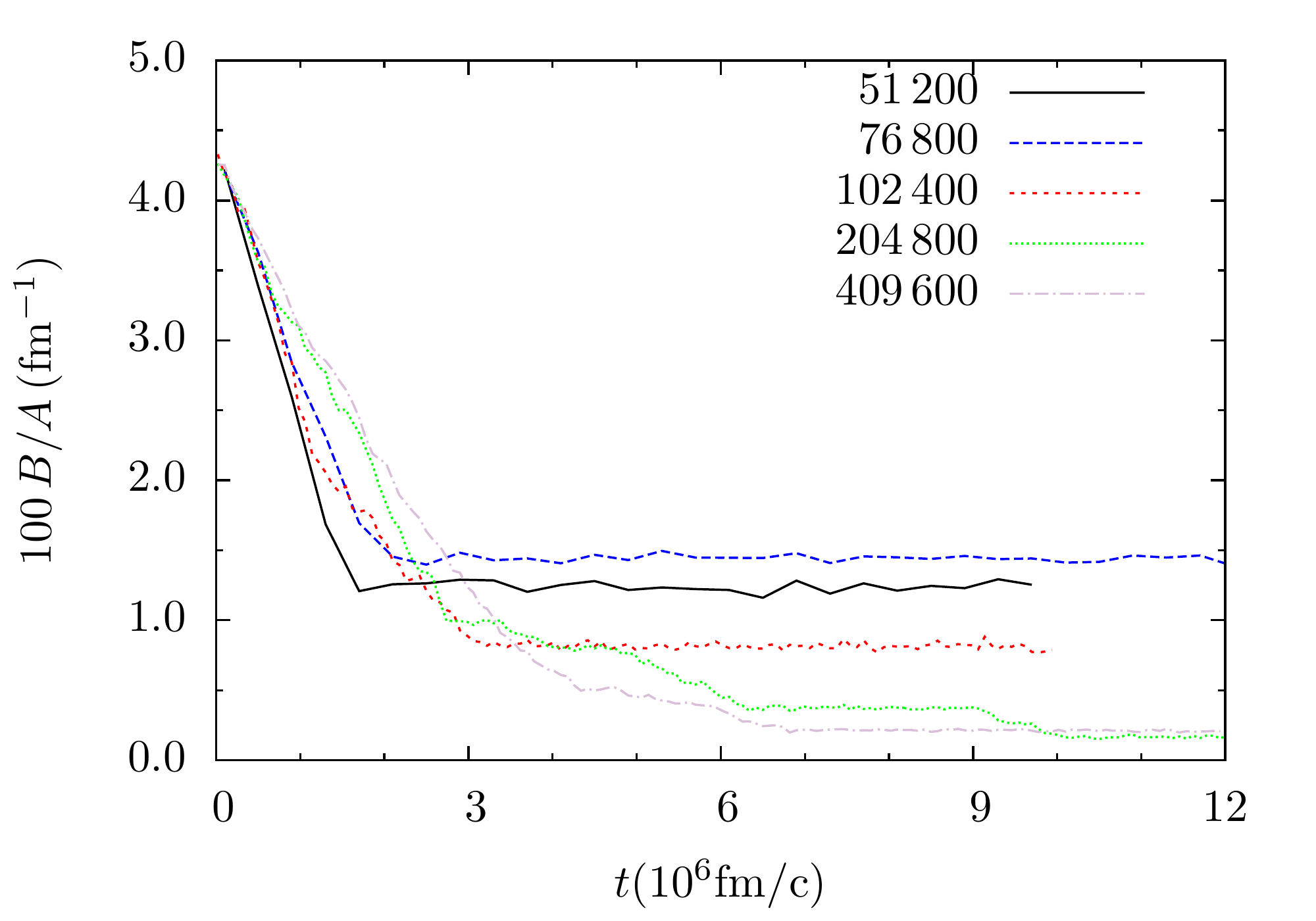}
\put (23,60) {(a)}
\end{overpic}
\begin{overpic}[width=0.50\textwidth]{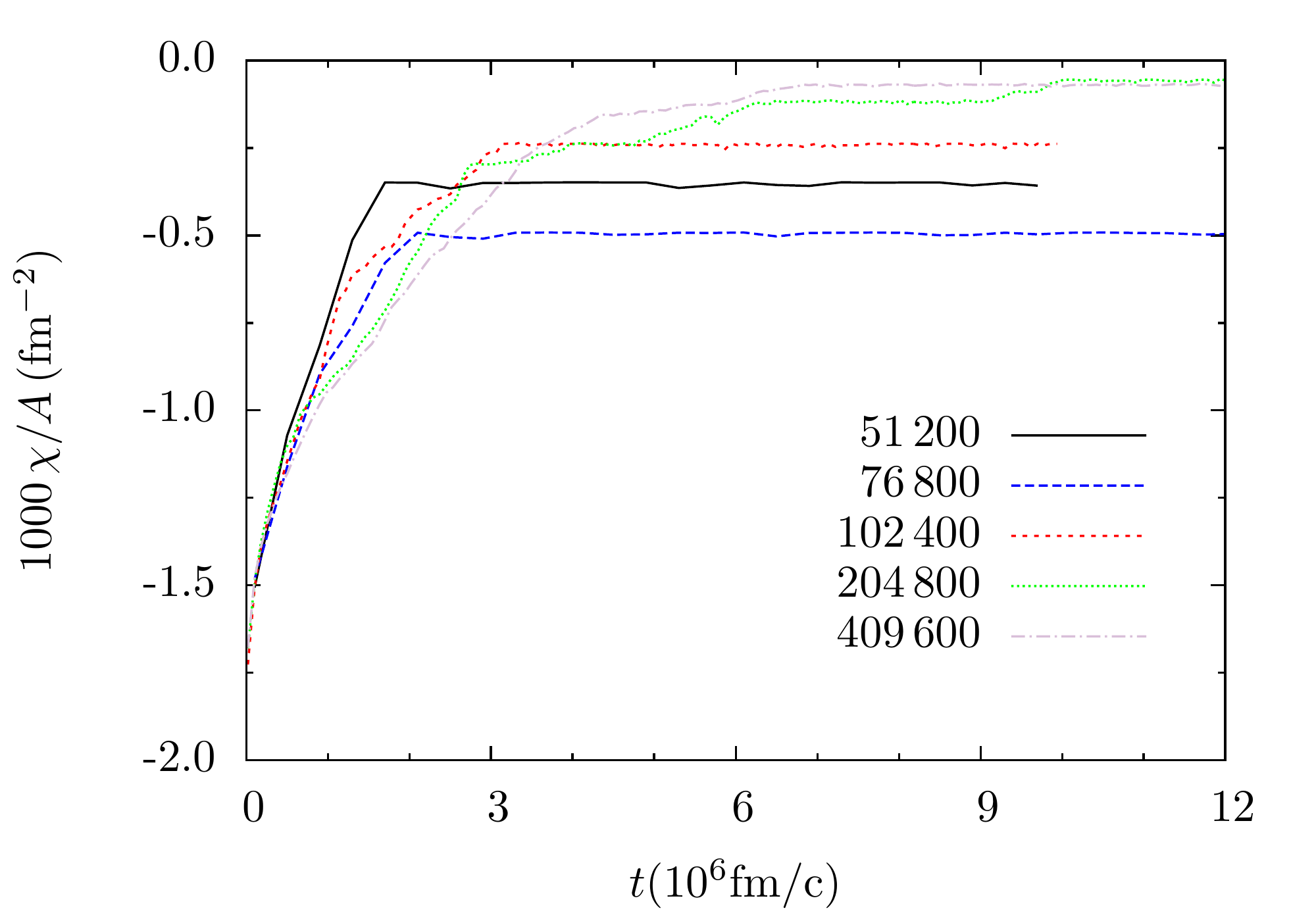}
\put (23,60) {(b)}
\end{overpic}
\caption{\label{fig:plots} (Color online) 
(a) Average mean curvature $B/A$ and (b) average Gaussian curvature $\chi/A$ as a function of time for impure pasta simulations. The curvatures for perfect pasta quickly reach values close to zero and, thus, not shown.}
\end{figure}

In Figure \ref{fig:plots} we plot the curvature of each simulation as a function of time. In the three smaller simulations both curvatures change close to linearly in the first 2 to 3 $10^6\unit{fm}/c$ of the simulation before reaching a steady state. The two larger runs, meanwhile, take longer to equilibrate with the 204\,800 nucleon simulation spending a significant amount of time in local equilibrium before quickly transitioning to an apparent steady state. Assuming the density of defects is proportional to the Gaussian curvature we do not observe any clear trend relating the density of defects and simulation size. In fact, the two larger simulations have very similar final curvatures. Though this may be an indication that finite size effects are well constrained for the two larger simulations even larger ones are needed to confirm that.

Besides simulating impure pasta we also perform simulations of perfect pasta. These are obtained for systems of the same five sizes described above and are shown in the last column of Figure \ref{fig:configurations}. To obtain configurations without defects we introduce an external sinusoidal potential with a $1\unit{MeV}$ amplitude and wave-number $2\pi N_p/L$ parallel to one of the sides of the simulation volume. Here $L$ is the side of the cubic box and $N_p$ is an integer number which corresponds to the number maxima/minima in the external potential and, consequently, the number of pasta sheets formed. 
We keep the external potential on for a relatively short time, $100\,000\unit{fm}/c$, over which its amplitude is decreased linearly from $1\unit{MeV}$ to zero. This proved enough time for perfect pasta to form.
After the removal of the external potential we let the system evolve for another $900\,000\unit{fm}/c$. In most cases defects form quickly after the complete removal of the external potential. In fact, for each simulation size we found only a single value of $N_p$ for which the parallel plates remained stable until the end of the $10^6\unit{fm}/c$ run. For these cases we let the plates evolve for another $2\times10^6\unit{fm}/c$ and in all cases the plates remained stable. In Table \ref{Tab:plates} we tabulate the values of $N_p$ which produced stable parallel plates and the average distance between the center of the plates as a function of simulation size. Clearly finite size effects still play a role in these simulations as, for example, the 76\,800 and 102\,400 were stable with the same number of plates despite their different sizes. Also, while the side $L$ of the cube of the 409\,600 nucleon run is a factor of two larger than the side of cube of the 51\,200 nucleon run the number of plates is proportionally smaller on the larger simulation.

\begin{table}[!h]
\caption{\label{Tab:plates} Number of plates $N_p$ and distance between center of plates $d$ for simulations of perfect pasta plates.}
\begin{ruledtabular}
\begin{tabular}{*{3}{c}}
     Nucleons   &  $N_p$  & $d$ (fm) \\
\hline
  \,\,51\,200   &  \,\,6  &   16.8   \\
  \,\,76\,800   &  \,\,7  &   16.5   \\
     102\,400   &  \,\,7  &   18.1   \\
     204\,800   &     10  &   16.0   \\
     409\,600   &     11  &   18.3   \\
\end{tabular}
\end{ruledtabular}
\end{table}

\subsection{Structure Factors}\label{ssec:Structure Factors}

In this section we compare the structure factors $S_i(\boldsymbol{q})$ for protons ($i=p$) and neutrons ($i=n$) for momentum transfers $\boldsymbol{q}=2\pi/L\,(l,m,n)$. Again, $L$ is the side of the cubic simulation volume while $l$, $m$ and $n$ are integers. As the density of vectors $\boldsymbol{q}$ increases with $L$ we will focus our analysis on the two runs with 409\,600 nucleons. 

To describe how the anisotropy in the pasta shapes affects the structure factor we first determine which $\boldsymbol{q}'$ is perpendicular to the bulk of plates formed in the simulation volume. Though we expect $\boldsymbol{q}'$ to coincide with the momentum transfer $\boldsymbol{q}$ that produces the maximum value in $S_i(\boldsymbol{q})$ we will see below that this may not be the case.

In our simulation of perfect pasta with 11 plates aligned parallel to the $xy$ plane, it is clear that $\boldsymbol{q}'=\pm2\pi/L(0,0,11)$. Meanwhile, the first group of maxima in $S_i(\boldsymbol{q})$ occurs for $\boldsymbol{q}_1=\pm2\pi/L\,(0,1,\pm11)$ and have magnitudes $S_p(\boldsymbol{q}_1)\sim650$ and $S_n(\boldsymbol{q}_1)\sim800$. The second group of maxima occurs for $\boldsymbol{q}_2=\pm2\pi/L\,(1,0,\pm11)$ and their magnitudes are $S_p(\boldsymbol{q}_2)\sim500$ and $S_n(\boldsymbol{q}_2)\sim610$. In both cases $S_i(\boldsymbol{q})\sim\langle\rho_i^*(\boldsymbol{q},t)\rho_i(\boldsymbol{q},t)\rangle_t$ as $\langle\rho_i(\boldsymbol{q},t)\rangle_t\sim0$, see Equation \eqref{eq:sq}. As mentioned above, due to the alignment of the sheets in the simulation, we expected the maxima in $S_i(\boldsymbol{q})$ to occur for $\boldsymbol{q}'=\pm2\pi/L\,(0,0,11)$. Yet this is not the case, the reason being that for $\boldsymbol{q}=\boldsymbol{q}'$ both right hand side terms in Equation \eqref{eq:sq} have very similar magnitudes of order $\mathcal{O}(10^5)$ leading to $S_i(\boldsymbol{q})\sim\mathcal{O}(1)$.

For the 409\,600 nucleons simulation of impure pasta the maxima in $S_i(\boldsymbol{q})$ occur for $\boldsymbol{q}_1=\pm2\pi/L\,(8,2,-7)$. Also, the magnitude of these maxima are much larger than for the perfect pasta with $S_p(\boldsymbol{q}_1)\sim25\,000$ and $S_n(\boldsymbol{q}_1)\sim31\,000$. Here the calculated first term on the right hand side of Equation \eqref{eq:sq} is about 50\% larger than the second term and there are no other maxima of similar magnitude. Thus, we infer $\boldsymbol{q}'=\boldsymbol{q}_1=\pm2\pi/L\,(8,2,-7)$ to be the direction perpendicular to the bulk of plates formed in the simulation.

\begin{figure}[!h]
\centering
\begin{overpic}[width=0.5\textwidth]{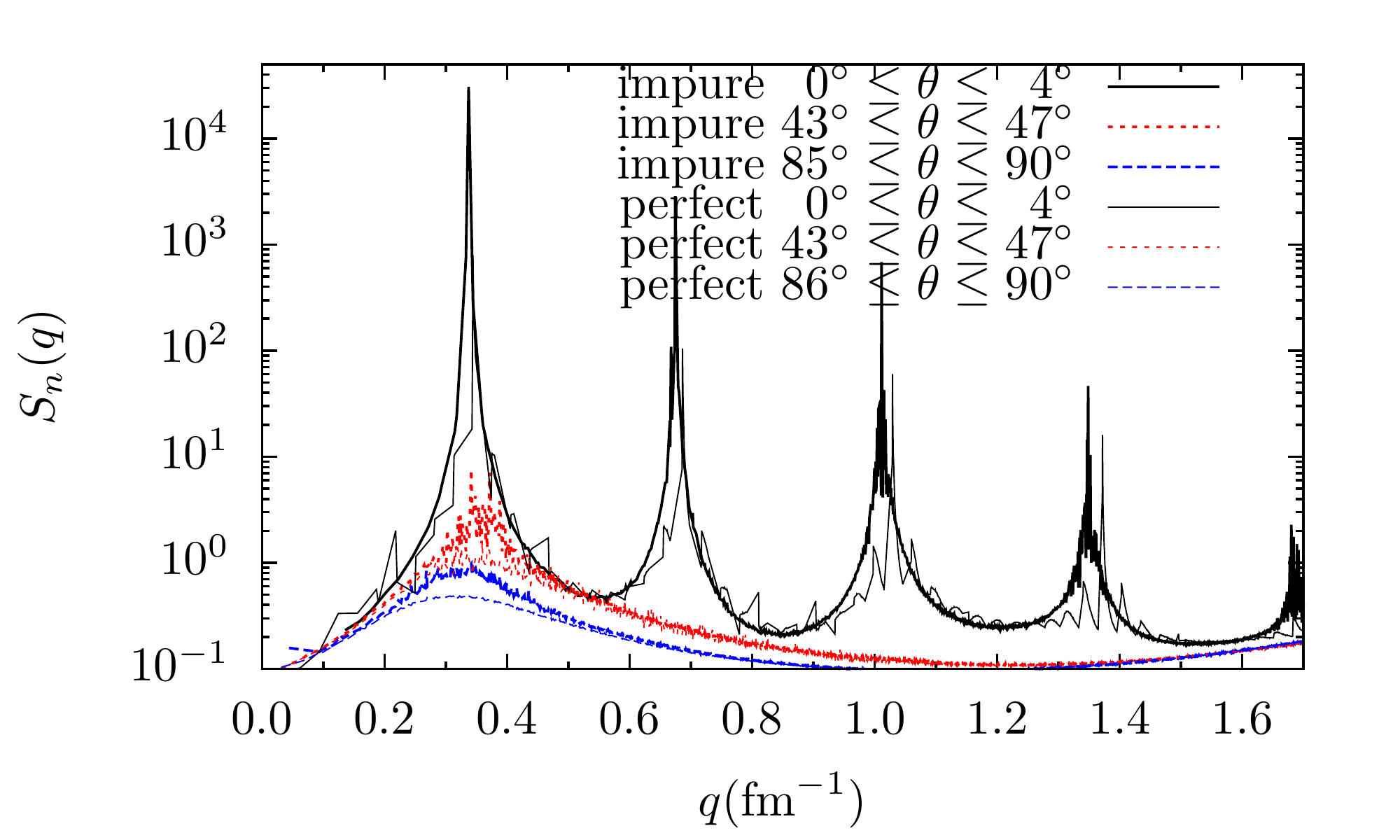}
\put (23,50) {(a)}
\end{overpic}
\begin{overpic}[width=0.5\textwidth]{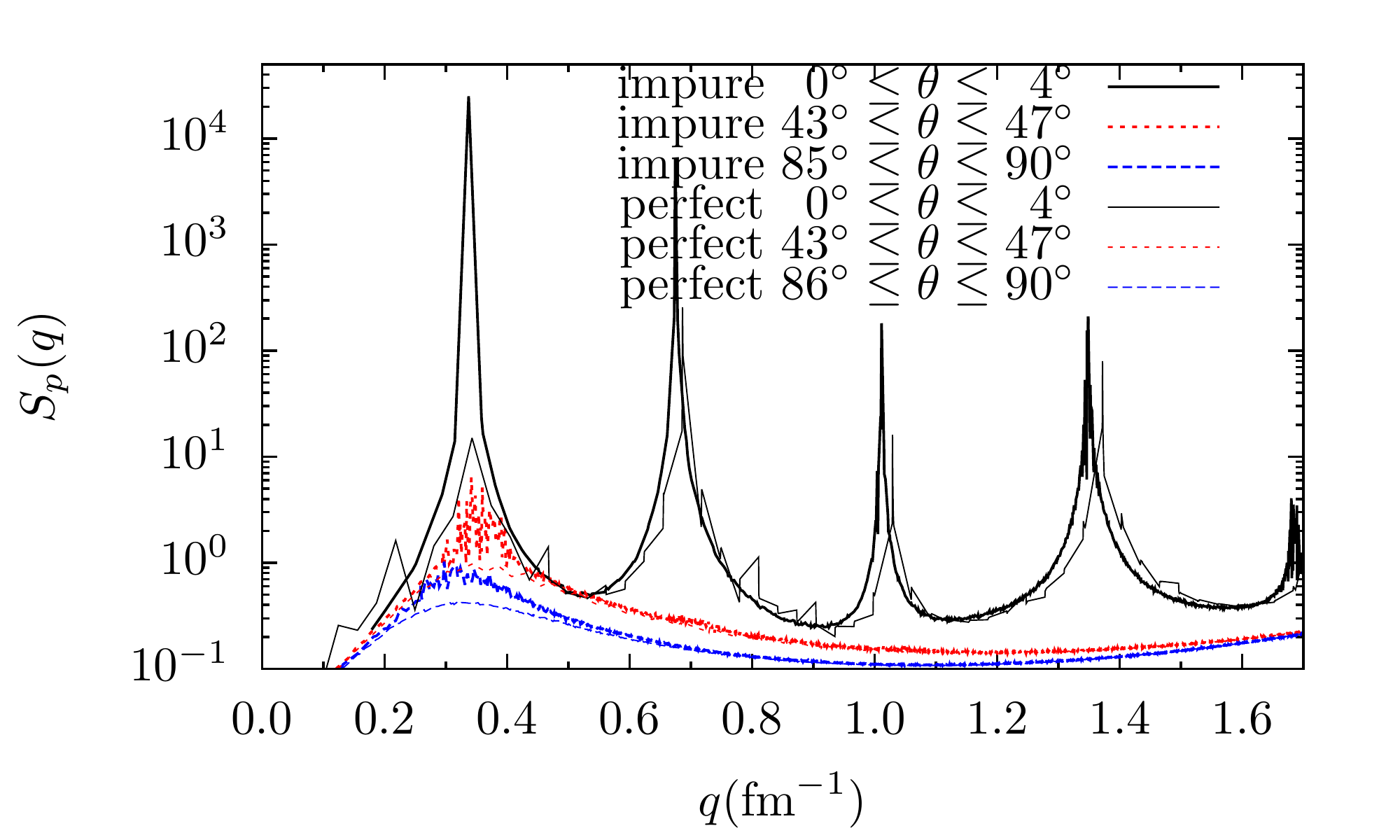}
\put (23,50) {(b)}
\end{overpic}
\caption{\label{fig:sq_anis} (Color online) 
Comparison of structure factor $S(\boldsymbol{q})$ for (a) neutrons and (b) protons for $\theta$ in three ranges: $0\degree\leq\theta\leq4\degree$ (solid black lines), $43\degree\leq\theta\leq47\degree$ (red short dashed lines) and $86\degree\leq\theta\leq90\degree$ (blue long dashed lines). Thin lines are for simulation of perfect plates while thick lines are for simulations of impure pasta.}
\end{figure}

In Figure \ref{fig:sq_anis} we plot $S_n(\boldsymbol{q})$ and $S_p(\boldsymbol{q})$ for vectors $\boldsymbol{q}$ that form some angle $\theta$ with $\boldsymbol{q}'$. The angle $\theta$ is defined by
\begin{equation}\label{eq:theta}
 \cos\theta=\frac{\boldsymbol{q}\cdot\boldsymbol{q}'}{\vert\boldsymbol{q}\vert\vert\boldsymbol{q}'\vert}.
\end{equation}
We limit the plots to the region $q=0$ to $2k_F$ as this is the region integrated to obtain the Coulomb logarithms, Equations \eqref{eq:visc2} and \eqref{eq:heat2}. We plot the structure factors $S_i(\boldsymbol{q})$ for three ranges of $\theta$: one where $\boldsymbol{q}$ is almost parallel to $\boldsymbol{q}'$ ($0\degree\leq\theta\leq4\degree$), one where $\boldsymbol{q}$ is almost perpendicular to $\boldsymbol{q}'$ ($86\degree\leq\theta\leq90\degree$) and, finally, one where $\boldsymbol{q}$ forms an angle of approximately 45\degree with $\boldsymbol{q}'$ ($43\degree\leq\theta\leq47\degree$).

First, we note in both plots of Figure \ref{fig:sq_anis} that the peaks in $S_i(\boldsymbol{q})$ occur periodically and decrease by orders of magnitude as $\boldsymbol{q}$ changes from being parallel to perpendicular to $\boldsymbol{q}'$. This is expected as the long range correlations occur for the pasta plates but not for the nucleons within the plates. Moreover, by comparing the results for the pasta with (impure) and without (perfect) defects at $\theta\sim0$ we observe that the peaks are wider and higher by an order of magnitude or more in the run that has defects. This may be due to defects decreasing the amplitude of low energy oscillations of the plates during a simulation. Lastly, we note that for the non parallel directions shown in the plots of Figure \ref{fig:sq_anis} the only significant differences in the structure factors $S_i(\boldsymbol{q})$ occur in the region of small momentum transfer, $q<0.5\unit{fm}^{-1}$.

\begin{figure}[!h]
\centering
\begin{overpic}[trim= 7 0 0 0,clip,width=0.5\textwidth]{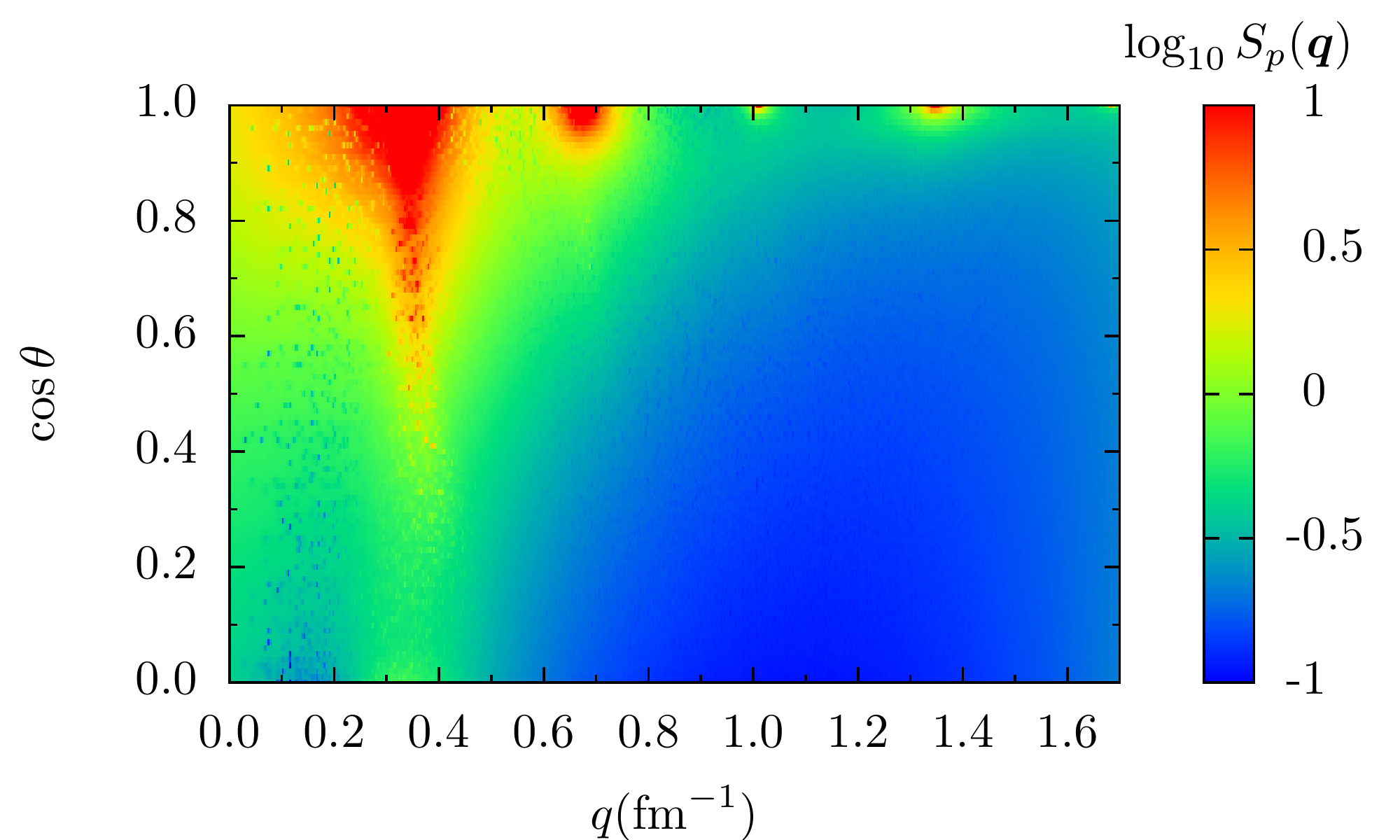}
\put (2,52) {(a)}
\end{overpic}
\begin{overpic}[trim= 7 0 0 0,clip,width=0.5\textwidth]{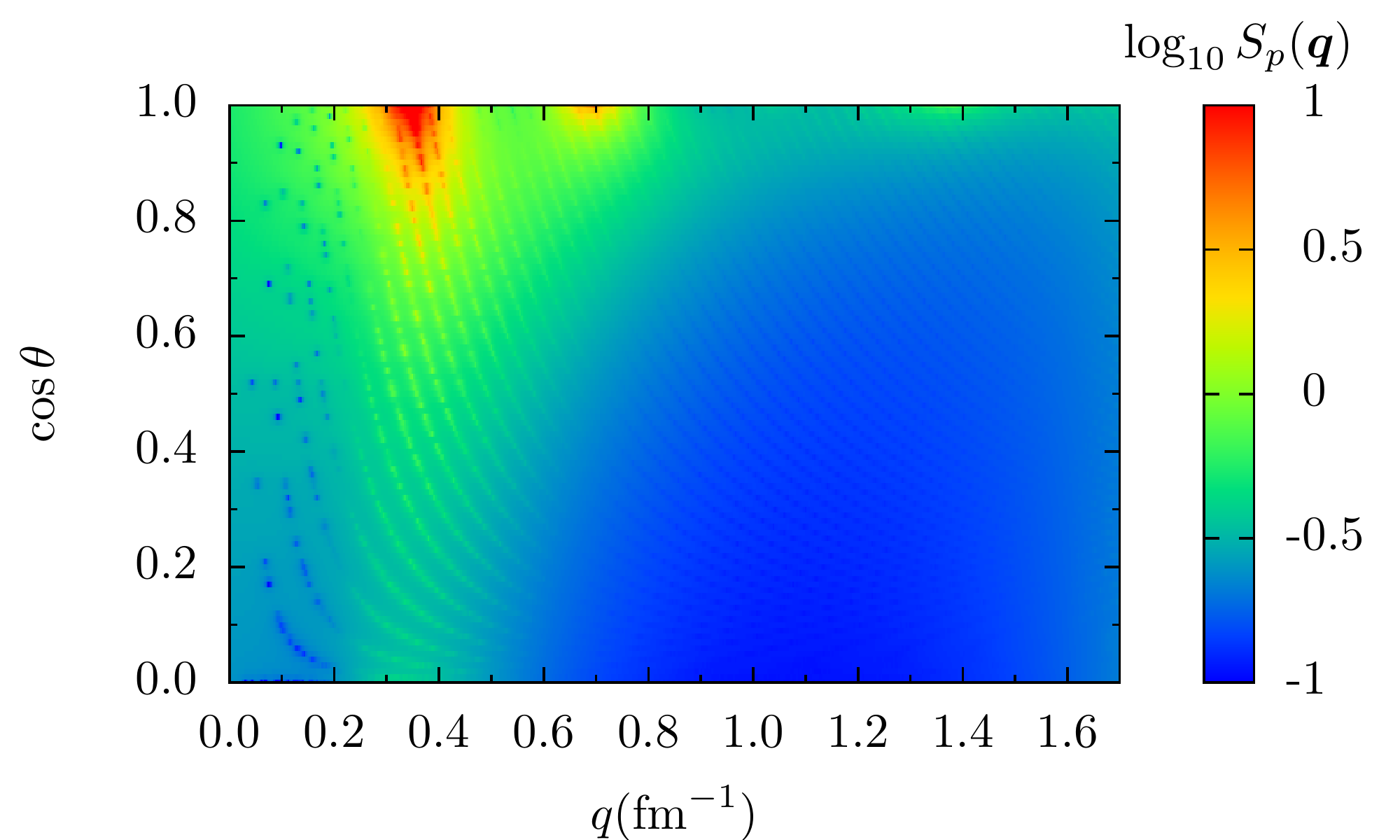}
\put (2,52) {(b)}
\end{overpic}
\caption{\label{fig:sq_full} (Color online) 
Heat map of proton structure factor $S_p(\boldsymbol{q})$  as a function of momentum transfer $q=\vert\boldsymbol{q}\vert$ and angle $\theta$ given by Equation \eqref{eq:theta} for pasta configurations (a) with defects and (b) without defects. Values of $\log_{10}S_p(\boldsymbol{q})$ were limited to the range -1 to 1 as very few points lie outside this range.}
\end{figure}

While in Figure \ref{fig:sq_anis} we focused on three ranges of $\theta$ ($\sim0\degree$, $\sim45\degree$ and $\sim90\degree$), in Figure \ref{fig:sq_full} we plot the results for the proton structure factor 
$S_p(\boldsymbol{q})$ for all $x=\cos\theta$ within 0 and 1 and $q$ within 0 and $2k_F$, see Equation \eqref{eq:theta}. Due to the finite size of the simulation only some values of $(q,x)$ are possible.
Thus, to obtain the plots in Figure \ref{fig:sq_full} we interpolate the values of $S_p(\boldsymbol{q})$ obtained from Equation \eqref{eq:sq} using the following equation 
\begin{equation}\label{eq:interpolation}
 S_p(q,x) = \frac{1}{\mathcal{N}}\sum_{\boldsymbol{k}} \frac{S_p(k,y)}{(k-q)^2+\alpha(x-y)^2}.
\end{equation}
Here $y=\cos\theta'$ is obtained from Equation \eqref{eq:theta} replacing $\boldsymbol{q}$ by $\boldsymbol{k}$ and the normalization $\mathcal{N}$ is given by
\begin{equation}
 \mathcal{N} = \sum_{\boldsymbol{k}} \frac{1}{(k-q)^2+\alpha(x-y)^2}
\end{equation}
where we set $\alpha=1\unit{fm}^{-1}$. We limit the range of the plot of $S_p(\boldsymbol{q})$ from $10^{-1}$ to $10^1$ as only a few values lie outside this range. The qualitative behavior for the neutron structure factor $S_n(\boldsymbol{q})$ is very similar to $S_p(\boldsymbol{q})$ and, thus, not shown.

The interpolation method of Equation \eqref{eq:interpolation} has problems, specially for regions near delta function like spikes. It introduces errors to the width of the peaks that should be determined. We postpone this to a future study where we intend to analyze even larger simulations and whether the width and height of the peaks changes as a function of simulation time. 

Both Figures \ref{fig:sq_anis} and \ref{fig:sq_full} show that the structure factor of perfect and impure pasta peak near $\cos\theta\sim1$, \textit{i.e.}, for $\boldsymbol{q}$ parallel to $\boldsymbol{q}'$, and that the peaks in $S_i(\boldsymbol{q})$ are higher for the impure pasta. Also, the maxima in $S_i(\boldsymbol{q})$ for all angles $\theta$ occur near the region $q\sim0.35\unit{fm}^{-1}$. This is because the strongest correlations on nucleon position occurs for adjacent pasta plates which are separated by approximately $2\pi/q\sim18\unit{fm}$.

\subsection{Viscosity and heat conductivity}\label{ssec:heat_visc}

In the pasta systems considered in this work only electron-nucleon scattering is expected to contribute significantly to the viscosity $\eta$ and heat (electronic) conductivity $\kappa$ ($\sigma$). Additionally, since the temperature is much smaller than the Fermi temperature, $T\ll T_F$, the contribution is mainly from electrons scattered in the Fermi surface, \textit{i.e.} with momentum $k_F$. For the electron densities considered here we have that the Fermi temperature and momentum are, respectively, $T_F=166\unit{MeV}$ and $k_F=0.84\unit{fm}^{-1}$. Since the electrons are scattered by the contrast in charge density it is expected that they will be scattered much more frequently when their momentum is perpendicular to the plates than when it is parallel \cite{PhysRevC.78.035806,Yakovlev11102015}. This is seen in the results for the structure factor $S_p(\boldsymbol{q})$ outlined in the previous section. For further discussion on electron-pasta collision frequencies including cases with magnetic fields see Reference \cite{Yakovlev11102015}.

\begin{figure}[!h]
\centering
\begin{overpic}[width=0.5\textwidth]{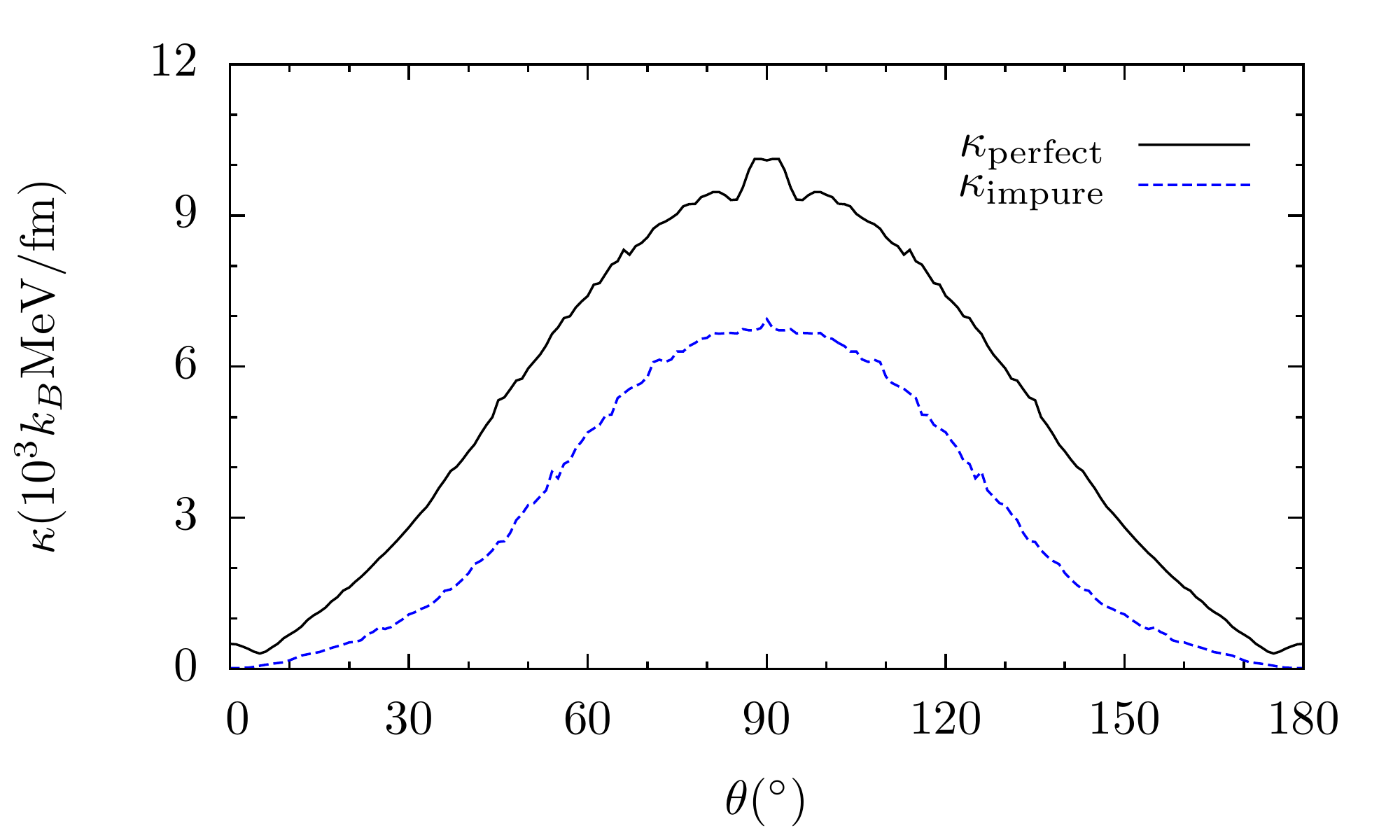}
\put (2,52) {(a)}
\end{overpic}
\begin{overpic}[width=0.5\textwidth]{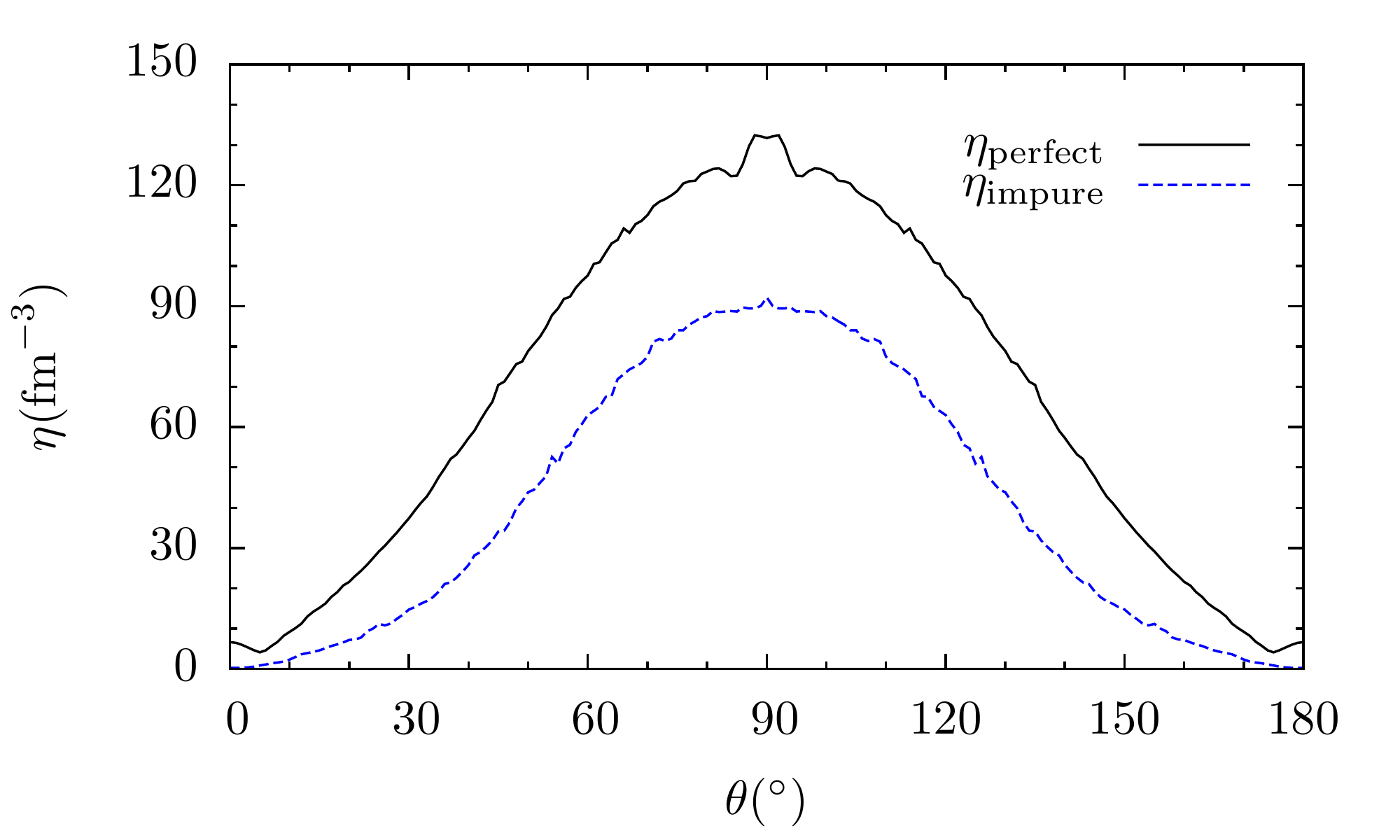}
\put (2,52) {(b)}
\end{overpic}
\caption{\label{fig:eta_kappa_z} (Color online) 
(Color online) Comparison of $(a)$ heat conductivity $\kappa$ and $(b)$ viscosity $\eta$ for the 409\,600 nucleon simulations for the pasta phases with defects (perfect pasta, dashed blue lines) and without defects (impure pasta, solid black lines) averaged over directions that form an angle $\theta$ with the vector $\boldsymbol{q}'$ perpendicular to the pasta slabs.}
\end{figure}

From the proton structure factor $S_p(\boldsymbol{q})$ for perfect and impure pasta we may estimate the Coulomb logarithms across different scattering directions. We use the interpolation method of Equation \eqref{eq:interpolation} to obtain $S_p(\boldsymbol{q})$ for different scattering angles $\theta$ and then integrate Equations \eqref{eq:visc2} and \eqref{eq:heat2} using a simple trapezoid method. The results obtained for the Coulomb logarithms are then used to obtain the viscosity and heat conductivity of the pasta. Since both Coulomb logarithms integrals are always within 5\% of each other the patterns for the heat conductivity $\kappa$ and viscosity $\eta$ are very similar. This is shown in the plots of Figure \ref{fig:eta_kappa_z}.

Now we discuss the heat conductivity and recall that similar conclusions may be drawn for both the viscosity and electrical conductivity of the pasta. As expected, heat conductivity peaks on directions nearly parallel to the plates, where charge density contrast is minimum, and plunges for directions nearly perpendicular to them, where charge contrast reaches its maximum. Thus, heat transport is much less efficient along directions normal to the slabs. In Section 2.2 of Reference \cite{Yakovlev11102015} Yakovlev discusses the relaxation timescales for electron scattering at angles parallel and perpendicular to pasta plates in the zero magnetic field limit for pasta formed of parallel pasta plates and other configurations. His conclusions are in agreement with our calculations. 

Our main goal was to determine how the presence of topological defects changes the transport properties of the pasta. The plots in Figure \ref{fig:eta_kappa_z} show there is a significant reduction in heat conductivity across all directions. The reduction is sharper along directions normal to the plates than parallel to them: the thermal conductivity decreases by a factor of 5 to 20 for small angles, $0\le\theta\le10\degree$, but only by about $30\%$ for larger angles, $60\lesssim\theta\le90\degree$. This result was expected as discussed in Reference \cite{PhysRevLett.114.031102}.

In Table \ref{Tab:eta_kappa} we write the average heat conductivity $\bar{\kappa}$, viscosity $\bar{\eta}$ and effective charge $\bar{Z}^*$. The angular average for an observable $O$ are defined by
\begin{equation}\label{eq:avg}
 \bar{O}=\frac{\int O(\theta)\sin\theta d\theta}{\int \sin\theta d\theta}.
\end{equation}
On average the defects decrease both heat conductivity and viscosity by 37\%. However, this effect is not uniform; it is much more pronounced for directions perpendicular to the slabs than parallel to them. Note that both average values are about one order of magnitude larger than the ones obtained for the $Y_p=0.20$ simulations of Horowitz and Berry in Reference \cite{PhysRevC.78.035806}. 

\begin{table}[!htb]
\caption{\label{Tab:eta_kappa} Average viscosity $\bar{\eta}$, heat conductivity $\bar{\kappa}$ and effective $\bar{Z}^*$ for the pasta with and without topological defects. Unit conversion factors are $1\unit{fm}^{-3}=10^{11}\unit{Pa}/s$ and $1k_B\unit{MeV/fm}=2\times10^{18}\unit{erg}/(\unit{K\,s\,cm})$.}
\begin{ruledtabular}
\begin{tabular}{*{5}{c}}
     Simulation   &  $\bar{\eta}(\unit{fm}^{-3})$  & $\bar{\kappa}(10^3k_B\unit{MeV/fm})$ & $\bar{Z}^*$ & \\
\hline
  perfect   & 87.7 &   6.66  & \,\,5.5 &  \\
  defects   & 55.5 &   4.15  &    50.2 &  \\
\end{tabular}
\end{ruledtabular}
\end{table}

If we assume that at the crust-core boundary of a neutron star there is a pasta phase such as the one described in this work and, furthermore, the bulk of the plates are aligned perpendicular to the radius of the star then most heat transfered from the core to the surface will be damped by the slabs of pasta. On the other hand, if the plates are parallel to the radius of the star then very little damping occurs as heat is transferred from the core to the crust. Still, in both cases, the presence of topological defects decreases the heat conductivity. If the pasta plates are randomly oriented then the defects would decrease the conductivity by 37\%. Nonetheless, this estimate depends on the density of defects being well estimated by our 409\,600 nucleon runs. This needs even larger simulations to be confirmed as that finite size effects could still be responsible for an underestimate of the defect density. For example, multiple defects separated by a distance larger than our simulation volume could be aligned in different directions. We expect this to decrease even further the viscosity and thermal conductivity of the system. Finally, the transport properties will be sensitive to the temperature and proton fractions in the inner crust of the NS which are expected to be much lower than  the $kT=1\unit{MeV}$ and $Y_p=0.40$ of our simulations.

\subsection{Impurity Parameter}\label{ssec:Impurity}

In Reference \cite{1993ApJ...404..268I} Itoh and Kohyama laid out a simple equation to determine the thermal conductivity of an ion systems contaminated by weakly-correlated impurities. They showed that the impurity contribution to the viscosity and heat conductivity was inversely proportional to the charge dispersion square of the ions, often referred to as the impurity parameter $Q_{\text{imp}}$:
\begin{equation}
 Q_{\text{imp}}=\langle(\Delta Z)^2\rangle=\langle{Z^2}\rangle-\langle{Z}\rangle^2.
\end{equation}
Meanwhile, Horowitz \textit{et. al.} performed molecular dynamics (MD) simulations to study how correlations amongst impurities affected the thermal conductivity of multicomponent plasmas. They showed that strong correlations between impurities played a significant role in the transport properties of the system \cite{PhysRevE.79.026103}. Daligault and Gupta also used MD simulations and showed that the impurity parameter formalist may be inadequate to describe systems with large $Q_{\text{imp}}$ and the thermal conductivity may deviate significantly from $\kappa\propto Q_{\text{imp}}^{-1}$ \cite{0004-637X-703-1-994}. Still, in this Section we obtain an estimate of $Q_{\text{imp}}$ for the system studied this work. 

For temperatures and densities considered in this work the heat transfer is dominated by electron-proton collisions and, thus, the thermal conductivity of the system can be written as \cite{0004-637X-703-1-994,0004-637X-698-2-1020}
\begin{equation}\label{eq:kappa1}
 \kappa=\frac{\pi^2}{3}\frac{k_B^2Tn_e}{m_e^*\nu^\kappa_{ep}}.
\end{equation}
Here $m_e^*$ is the electron effective mass and $n_e$ the electron density. When working in the nucleon degrees of freedom and an electrically neutral system, the electron-proton collision frequency $\nu^\kappa_{ep}$ is given by
\begin{equation}\label{eq:freq}
 \nu^\kappa_{ep}=\frac{4\pi\alpha^2n_e}{k_F^2v_F}\Lambda^\kappa_{ep}.
\end{equation}
As in Section \ref{sec:Formalism} $k_F$ and $v_F$ are the Fermi momentum and velocity while $\Lambda^\kappa_{ep}$ is the Coulomb logarithm and encodes the structure of the system. Since the electrons are relativistic and highly degenerate $v_F\sim1$, $m_e^*\sim k_F$ and we may substitute Equation \eqref{eq:freq} in \eqref{eq:kappa1} to obtain \eqref{eq:heat1}.

We may write the electron-proton collision frequency $\nu$ for the system with topological defects as
\begin{equation}
 \nu = \nu_0 + \nu_{\text{imp}}
\end{equation}
where $\nu_0$ ($\nu_{\text{imp}}$) is the contribution of the slabs (defects) to the collision frequency. Following the works of Brown and Cumming \cite{0004-637X-698-2-1020} and Daligault and Gupta \cite{0004-637X-703-1-994} the impurity contribution may be written as
\begin{equation}\label{eq:freqimp}
 \nu_{\text{imp}}=\frac{4\pi Q_{\text{imp}}\alpha^2n_{\text{pasta}}}{k_F^2v_F}\Lambda^\kappa_{\text{imp}}
\end{equation}
where $Q_{\text{imp}}$ is the impurity parameter, $\Lambda^\kappa_{\text{imp}}$ the impurity Coulomb logarithm and $n_{\text{pasta}}$ the density of pasta clusters. The pasta cluster density is related to the electron density by $n_e=\langle{Z}\rangle n_{\text{pasta}}$ where $\langle{Z}\rangle$ is the average charge of each pasta cluster. Note that for an ion system $n_{\text{pasta}}$ is equivalent to the ion density while $\langle{Z}\rangle$ the average ion charge.

Following the discussion in Appendix A1 of Brown and Cumming we determine the average charge in the inner crust to be $\langle{Z}\rangle=27.88$ \cite{0004-637X-698-2-1020}. Furthermore, for the system analyzed here we have $\Lambda^\kappa_{\text{imp}}=2.06$, see Equation (A6) from Brown and Cumming, Reference \cite{0004-637X-698-2-1020}, adapted from Potekhin \etal\,\cite{1999A&A...346..345P}. Though this formalism assumes that the impurities are uncorrelated, which is not the case for the topological defects connecting the slabs of pasta, we still use it to provide an estimate for the impurity parameter $Q_{\text{imp}}$.
With some algebra we obtain that
\begin{equation}\label{eq:qimp}
 Q_{\text{imp}} = \frac{\langle{Z}\rangle}{\Lambda^\kappa_{\text{imp}}}\left(\bar{\Lambda}^\kappa_{\text{impure}}-\bar{\Lambda}^\kappa_{\text{perfect}}\right).
\end{equation}
As above, the bars in $\bar{\Lambda}^\kappa_{\text{impure}}$ and $\bar{\Lambda}^\kappa_{\text{perfect}}$ denote angle averaged quantities. Assuming that the pasta in the crust-core boundary and its defects are randomly oriented an angular average of $Q_{\text{imp}}$ should give a good description of the impurity parameter due to the defects. Thus, using the values quoted above we obtain $\bar{Q}_{\text{imp}}=29$ a value that closely agrees with the value $Q_{\text{imp}}$ between 30 to 40 estimated in Reference \cite{PhysRevLett.114.031102}. This value is, however, smaller than the value of $Q_{\text{imp}}$ used by Pons \etal\, in Reference \cite{pons2013highly}.

\section{Conclusions}\label{sec:Conclusions}

Using large scale molecular dynamics (MD) simulations we compared the structure and observables of nuclear pasta formed by perfect parallel plates (perfect pasta) and by parallel plates connected by long lived topological defects (impure pasta) \cite{PhysRevLett.114.031102,Horowitz:2015gda}. We quantified the anisotropy in the structures obtaining the structure factors along different directions and calculating transport properties of the systems. We showed that topological defects in the pasta phase of nuclear matter play a role similar to that of impurities in ion plasmas by reducing the thermal and electrical conductivities as well as the viscosity of the system. On average viscosity and thermal conductivity were reduced by 37\% due to the presence of defects connecting the plates. This effect on the observables was much more pronounced on directions perpendicular to the slabs than parallel to them.

In a previous work, using observations of the light curve of cooling LMXB 1659-29, we estimated a pasta impurity parameter within the neutron star crust of order $Q_{\text{imp}}\sim30$ \cite{PhysRevLett.114.031102}. 
Here we estimated the impurity parameter $Q_{\text{imp}}$ due to topological defects connecting pasta plates by comparing its observables to the ones of a system of plates without defects. From a procedure similar to the one used to determine the impurity parameter of ion systems \cite{0004-637X-703-1-994,0004-637X-698-2-1020,1999A&A...346..345P} we estimated an impurity parameter $Q_{\text{imp}}\sim29$ for slabs of nuclear pasta connected by topological defects. This estimate is in agreement with our simple estimate in a previous work \cite{PhysRevLett.114.031102} and in rough agreement with the value used by Pons \etal\, for the inner crust of a neutron star \cite{pons2013highly}. 
It is worth noting that we did this while ignoring the fact that the anisotropy of the system coupled to strong long range correlations of the defects could significantly alter this value. Furthermore, we studied systems with temperatures and proton fractions much larger than the ones expected in the inner rust of neutron stars. Yet, as of now we know of no other direct calculations of impurity parameters for pasta phase due to topological defects. 

For future works it may be important to obtain impurity parameters due to other types of topological defects, realistic temperatures and proton fractions and in the presence of magnetic fields as those are relevant for the physics of neutron stars.

\begin{acknowledgments}

A.\,S.\,S. would like to thank CNPq for the financial support through the Ci\^encia sem Fronteiras fellowship. We would like to thank Indiana University for access to the BigRed II supercomputer. This research was supported in part by DOE grants DE-FG02-87ER40365 (Indiana University) and DE-SC0008808 (NUCLEI SciDAC Collaboration). 

\end{acknowledgments}


\bibliography{s_of_q}

\end{document}